\documentclass[manuscript]{acmart}
\AtBeginDocument{%
  }

\usepackage{tcolorbox}
\tcbuselibrary{breakable}   % needed for the `breakable` option in your box
\usepackage{fancyvrb}
\usepackage{fvextra}

\newtcolorbox{answerbox}[1]{
    breakable,
    colback=white,
    colframe=gray!80,
    boxrule=0.5pt,
    arc=1mm,
    title=#1,
    fonttitle=\bfseries,
    left=2mm,
    right=2mm,
    top=1mm,
    bottom=1mm,
    toptitle=1mm,
    bottomtitle=1mm
}

\setcopyright{none}
\begin{document}

\title[The Changing Landscape of Online Knowledge Consumption and Production in the Age of Generative AI]{Navigating the Changing Landscape of Online Knowledge Consumption and Production in the Age of Generative AI: Evidence from Stack Overflow}

\author{Ji Eun Kim}
\email{jieunkim@umich.edu}
\orcid{0009-0008-1414-7341}
\affiliation{%
  \institution{University of Michigan}
  \city{Ann Arbor}
  \state{Michigan}
  \country{USA}
}

\author{Léa Vitale}
\email{vitalel@umich.edu}
\orcid{0009-0004-8944-8641}
\affiliation{%
  \institution{University of Michigan}
  \city{Ann Arbor}
  \state{Michigan}
  \country{USA}
}

\author{Libby Hemphill}
\email{libbyh@umich.edu}
\orcid{0000-0002-3793-7281}
\affiliation{%
  \institution{University of Michigan}
  \city{Ann Arbor}
  \state{Michigan}
  \country{USA}
}

\author{Yulin Yu}
\email{yulinyu@arizona.edu}
\orcid{0000-0003-4743-8360}
\affiliation{%
  \institution{University of Arizona}
  \city{Tucson}
  \state{Arizona}
  \country{USA}
}

\renewcommand{\shortauthors}{Kim et al.}

\begin{abstract}
  Online knowledge communities rely on a division of epistemic labor between users who seek information and those who produce it. Generative AI may blur these roles, but how it reallocates knowledge-seeking and knowledge-producing activities and reshapes the nature and returns of participation remains unclear. We examine changes in question-asking and answering among Stack Overflow user groups, focusing on consumers and producers, following the release of ChatGPT. Consumers shifted from asking questions to producing answers, but their answers were less likely to be accepted relative to the pre-ChatGPT period, suggesting that increased production did not yield equal standing in the community. By contrast, producers did not change the number of questions they asked, but increasingly asked about novel and emerging topics. Although they produced fewer answers, their answers received greater recognition. These findings show that access to generative AI does not necessarily translate into equal opportunities for successful participation.

\end{abstract}




\received{20 February 2007}
\received[revised]{12 March 2009}
\received[accepted]{5 June 2009}

\maketitle

\section{Introduction}
\label{intro}
Online knowledge-sharing platforms serve as accessible and democratized infrastructures for learning and problem solving, broadening access to expertise across geographic and temporal boundaries \cite{faraj2011knowledge}. On these platforms, users seeking knowledge pose questions, while users with relevant expertise provide answers. This exchange benefits both sides; askers receive tailored help that generic references cannot provide, and answerers refine their own skills and build reputation within the community \cite{bosu2013building, jin2015users}. As these exchanges accumulate, they form large, searchable repositories that benefit not only the original participants but also future knowledge seekers. The value and sustainability of these platforms therefore depend on continued participation by both those who seek information and those who provide it.

Recent advances in generative AI, however, may reshape this division of epistemic labor by changing both how users seek information and how they produce it. Tools such as ChatGPT can provide immediate, personalized answers outside community-based knowledge-sharing platforms. At the same time, these tools can help users draft responses and refine content, thereby lowering barriers to participation and complementing human knowledge production. Prior research provides evidence of both possibilities. On Wikipedia, page views and editing activity declined more for articles whose content substantially overlapped with ChatGPT-generated responses, suggesting that generative AI can draw attention and contributions away from existing knowledge repositories \cite{lyu2025wikipedia}. On Reddit, by contrast, newcomers increased their posting activity in subreddits centered on objective or factual topics, where generative AI may assist with content creation \cite{shorakaei2025empowering}. Generative AI can therefore function as both a substitute for and a complement to community participation. These mechanisms may operate differently even within the same platform because users' existing roles, expertise, and participation histories may shape whether AI reduces their reliance on the community, enables new forms of contribution, or alters how they move between seeking and producing knowledge.

Yet platform-level changes in engagement cannot reveal how generative AI redistributes participation across users. The same aggregate decline in contributions could reflect uniform disengagement across the community or offsetting behavioral changes among groups. This distinction matters because generative AI does not necessarily affect everyone uniformly \cite{capraro2024impact}. Prior research has identified distinct user roles in online communities \cite{saxena2022users, yang2016did}, but we know less about how generative AI may reconfigure the participation of these groups within the same community. Rather than asking only whether overall engagement increased or decreased, we examine how changes following the release of ChatGPT varied across Stack Overflow user segments with different prior participation roles.

We chose Stack Overflow as our study setting because it provides a well-defined setting for examining how generative AI reshapes the production and consumption of knowledge. Unlike platforms centered on advice, opinions, or personal experiences, Stack Overflow is organized around practical, well-scoped, and answerable programming problems, often accompanied by code and detailed technical context. Although programming questions may still admit multiple valid solutions, their correctness and usefulness are generally more externally verifiable, reducing variation in what constitutes a beneficial contribution. The platform also clearly distinguishes knowledge-seeking from knowledge-providing activities through questions and answers, while votes and accepted answers provide observable measures of community recognition and contribution outcomes. Moreover, because programming is a domain in which generative AI can directly substitute for or support both seeking and providing knowledge, Stack Overflow offers a theoretically relevant context for studying changes in these roles. Finally, its large-scale longitudinal records of questions, answers, and users enable systematic analyses of behavioral changes before and after the introduction of ChatGPT.

While previous studies have examined how overall trends in user activities changed following the launch of ChatGPT, they have not systematically examined how the impact of AI varies across user segments, particularly on Stack Overflow. We still know little about whether and how generative AI is associated with changes in the established division of knowledge labor within online Q\&A communities, where users occupy distinct roles in seeking and producing knowledge. We address this gap by examining behavioral changes across Stack Overflow user groups following the release of ChatGPT. We consider not only changes in question-asking and answer production but also the outcomes of participation.

\begin{itemize}
\item {\textbf{RQ1}}: How did the volume and distribution of question-asking and answer-providing activities across users with different preexisting roles change following the release of ChatGPT?

\item {\textbf{RQ2}}: How did the nature and returns of participation change across user roles following the release of ChatGPT?

\end{itemize}

Through interrupted time series analysis, we find substantial changes in how different user groups sought and produced programming knowledge on Stack Overflow following the release of ChatGPT. Consumers, who primarily asked questions before the release of ChatGPT, subsequently posted fewer questions but more answers. In contrast, producers showed no significant change in the number of questions they asked but contributed significantly fewer answers. Consumers’ answers showed a significant increase in similarity to AI-generated responses following the release of ChatGPT, providing evidence consistent with AI-assisted knowledge production. However, this expansion of participation did not narrow the gap between consumers and producers in the outcomes of their participation. Consumers’ share of answers accepted by question askers declined, whereas producers continued to maintain a higher share of their answers accepted and experienced a further increase after ChatGPT’s release. Additionally, producers were more likely to ask questions at the knowledge frontier, including questions involving novel and rapidly emerging topics. Together, these findings suggest that generative AI may enable users to move beyond their established roles without producing comparable benefits across groups; consumers contributed more answers, but producers remained better positioned to receive recognition for their answers and engage with frontier knowledge. By revealing these group-specific changes, this study contributes to a more nuanced understanding of how generative AI reshapes online knowledge communities and highlights the need to examine not only overall activity but also how changes in participation and their associated benefits vary across user groups.

\section{Related Work}

We situate our work within research on online knowledge communities as interdependent knowledge infrastructures sustained by complementary roles in seeking and producing information. We then review how generative AI can reshape both activities, functioning as a substitute for community participation while also supporting user contributions. Finally, we examine emerging research on how these effects vary across user groups, focusing on changes in participation and its outcomes.

\subsection{Knowledge Consumption and Production in Online Communities}

Although reading and searching are also forms of consumption \cite{sun2014understanding}, this study focuses on question-asking as an observable form of public knowledge consumption. We define knowledge production as answer posting. Online knowledge communities depend on the complementary functions of questions and answers. Questions articulate information needs, identify unresolved problems, and create opportunities for others to contribute expertise \cite{shi2021questions, bighash2018value, ravi2014great}. Answers address those needs and transform individual exchanges into reusable knowledge resources \cite{fichman2011comparative}. Question-asking and answer posting are therefore interdependent. Without questions, communities have fewer opportunities to generate new knowledge, while without answers, questions provide limited value to either askers or future visitors. 

Because questions and answers serve distinct but complementary functions, users naturally differ in the extent to which they engage in each activity \cite{saxena2022users}. Some primarily seek knowledge by asking questions (consumers), whereas others primarily produce knowledge by providing answers (producers). Hybrid users regularly engage in both activities, while low-activity users contribute too infrequently to exhibit a stable participation pattern. These participation patterns may become stable because asking and answering require different resources and offer different returns. Producing answers often requires domain expertise, confidence, time, and familiarity with community norms, whereas asking questions generally requires users to recognize and articulate a problem they cannot resolve independently. Reputation systems may further reinforce specialization by rewarding users for activities in which they have already developed experience and status \cite{li2012quantifying, vasilescu2014social}. Consequently, users may continue occupying roles that match their capabilities, needs, and expected benefits. The sustainability of a shared knowledge repository nevertheless depends on participation across these roles; consumers generate the questions that motivate knowledge production, while producers supply and evaluate the answers that make the repository useful. Technologies that change the costs or capabilities associated with asking and answering may therefore alter not only overall activity but also the established distribution of knowledge-seeking and knowledge-producing roles \cite{tausczik2020knowledge}.

Generative AI, such as ChatGPT, has fundamentally altered how users engage with several prominent knowledge-sharing platforms, including Stack Overflow \cite{gallea2023mundane, helic2026stack, shan2025examining}, Wikipedia \cite{lyu2025wikipedia, reeves2025exploring, brooks2024rise, zhou2025llms}, Reddit \cite{shorakaei2025empowering}, and GitHub \cite{kreitmeir2024heterogeneous}. On Stack Overflow, prior studies have documented a marked decline in website visits and the volume of posts \cite{gallea2023mundane, burtch2024consequences, del2024large, helic2026stack}. In addition, both questions and answers have become longer and exhibit increased complexity \cite{helic2026stack}. Taken together, these findings suggest that activity on Stack Overflow has declined, while the platform has assumed a more specialized role as users increasingly turn to LLMs for less complex information \cite{xue2026can}. The decline in user engagement on Stack Overflow may be attributed to ChatGPT’s ability to provide immediate answers without waiting for community input, facilitate brainstorming, and offer personalized learning support \cite{garcia2023exploring}. The decline in user engagement may also reflect a self-reinforcing cycle; as users leave the platform, contributors receive less reciprocal help and social interaction, weakening their motivation to continue sharing knowledge \cite{mustafa2023motivates}. Nevertheless, Stack Overflow remains valuable for resolving specific and complex debugging problems through community expertise and for providing links to relevant resources \cite{liu2023chatgpt}. Thus, ChatGPT and Stack Overflow coexist as complementary programming tools, each offering distinct strengths. To sum up, online communities rely on differentiated but complementary participation roles, yet these roles are not necessarily fixed in the age of generative AI. The next subsection reviews how generative AI can change both knowledge seeking and production.

\subsection{Generative AI as a Substitute for and Support for Community Participation}

Generative AI can substitute for participation in online knowledge communities. Instead of posting questions publicly, users can obtain immediate and personalized responses from tools such as ChatGPT \cite{liu2023chatgpt, liang2024large}. This shift may reduce question volume and, in turn, the opportunities and incentives for established producers to provide answers, contributing to declines in platform visits, posts, and other forms of community activity. Unlike public question-and-answer exchanges, however, private interactions with AI are not scrutinized by community members or preserved in searchable archives. Thus, even when generative AI satisfies individuals’ immediate information needs, substituting private AI use for public participation may reduce the production of shared knowledge resources.

At the same time, generative AI can support participation by assisting users with drafting, revising, explaining, translating, and formatting contributions \cite{skjuve2024people, fui2023generative}. Such assistance may reduce the expertise, confidence, time, and effort required to formulate an answer that meets community expectations \cite{noy2023experimental}. Generative AI may therefore enable consumers and novice users who previously lacked the capability to answer questions to attempt knowledge production. It may also accelerate experienced producers’ work \cite{zhou2025llms}, allowing them to address problems more efficiently or exposing them to more advanced questions as they progress to unfamiliar tasks. To sum up, generative AI may consequently draw some activity away from knowledge-sharing platforms while enabling new forms of participation within them, with its effects depending on users’ prior roles and capabilities.

\subsection{Heterogeneous Effects of Generative AI Across User Groups}

Prior research has developed typologies of user roles across knowledge-sharing platforms to support task allocation and provide role-specific assistance \cite{saxena2022users, yang2016did}. With the rapid adoption of generative AI, scholars have increasingly examined how this technology changes the ways users seek information, contribute knowledge, and participate in online communities. However, these changes may not affect all groups equally because users differ in their expertise, experience, and established participation patterns.

Emerging evidence points to such heterogeneous effects across platforms. On Reddit, expert users reduced their activity in certain subreddits characterized by a high proportion of loyal users following the release of ChatGPT \cite{shorakaei2025empowering}. On GitHub, excessive reliance on ChatGPT may disproportionately disadvantage less-experienced software developers because errors in AI-generated code can make debugging more difficult \cite{kreitmeir2024heterogeneous}. Similarly, an interview study of Wikipedia editors found that experienced editors used generative AI to explore and contribute to new topics, whereas newcomers continued to face difficulties when their AI-assisted edits failed to conform to community guidelines and were rejected \cite{zhou2025llms}. On Stack Overflow, AI adopters tend to produce shorter and more readable answers, but that excessive reliance on AI could ultimately hinder productivity \cite{shan2025examining}. Collectively, these studies suggest that the effects of generative AI depend on users’ prior expertise, experience, and familiarity with community norms.

Although generative AI has shown promise for narrowing productivity gaps between high- and low-performing individuals across several domains \cite{noy2023experimental, brynjolfsson2025generative}, its benefits may remain unevenly distributed in other contexts and among different populations \cite{capraro2024impact}. Existing research therefore leaves two related questions unresolved: whether generative AI alters participation patterns associated with established user roles and whether users who begin engaging in activities outside their prior roles achieve outcomes comparable to those of more experienced participants. Addressing these questions is essential for understanding whether generative AI merely broadens participation or also distributes the benefits of AI-mediated participation more equitably.

\section{Data and Methods}
\label{data_methods}

\subsection{Data}
\label{data}
To answer our research questions, we retrieved Stack Overflow data from the Internet Archive.\footnote{\url{https://archive.org/download/stackexchange}} Our observation period spanned two years, from one year before the release of ChatGPT to one year after its release (November 30, 2021–November 29, 2023). The total numbers of questions and answers posted during our observation period are 1,806,827 and 2,479,528, respectively. We also collected information on 926,739 unique users who posted at least one question or answer during the pre-ChatGPT period. 

\subsection{Role Identification}
\label{role_ident}
We categorized users who were active during the pre-ChatGPT period into four groups: consumers, producers, hybrid users, and low-activity users. First, we applied a minimum activity threshold of 10 items, each of which could be either a question or an answer. This ensures that role assignment reflects meaningful activity levels, rather than a small number of actions that could make it noisy and unstable. Users who posted fewer than 10 total items in the pre-ChatGPT period were labeled as low-activity users, because we did not have enough historical data to define their role. For those who posted at least 10 items, we computed the proportions of questions and answers for each user using only pre-ChatGPT activity data. If a user had a proportion of questions of at least 80\%, we labeled them as a consumer, someone who mainly seeks answers to their questions from other users. If a user had a proportion of answers of at least 80\%, we labeled them as a producer, someone who mainly provides answers to others' questions. Everyone else was classified as a hybrid user, tending to both ask and answer questions. Stack Overflow exhibits a skewed distribution of user engagement, a pattern commonly observed in many online communities \cite{sun2014understanding}; about 95\% of users belong to the low-activity group, while 1.2\%, 2.2\%, and 1.2\% of users belong to the consumer, producer, and hybrid groups, respectively. A sensitivity analysis using alternative minimum-activity and role-assignment thresholds yielded substantively consistent results across all specifications (see Appendix \ref{sen_anal}).

\subsection{Interrupted Time Series Analysis}
\label{analysis}
This study uses several variables to assess how the extent, nature, and outcomes of user participation change over time. All outcome variables are aggregated to the group-week level for an interrupted time series (ITS) analysis \cite{morgan2014counterfactuals, bernal2017interrupted, linden2015conducting}, which allows us to examine how patterns differ before and after a major event. The design models historical trends to construct a counterfactual scenario in which the event did not occur, and then compares this counterfactual trend with the observed post-event trend to distinguish sudden shifts from gradual changes. Specifically, our ITS design models (1) the pre-event trend, (2) the immediate change at the event time, and (3) the post-event trend, to investigate how different user groups' activity patterns changed following the introduction of ChatGPT. 

We estimate the following ITS model: 
\begin{equation}
\begin{aligned}
Y_{gt} = \beta_0 + \beta_1 Time_t + \beta_2 Post_t + \beta_3 TimeAfter_t + \gamma_g + \phi_g Time_t + \delta_g Post_t + \theta_g TimeAfter_t + \epsilon_{gt},
\end{aligned}
\end{equation}
where one user group is designated as the reference group. Here, $g$ denotes the user group and $Time_{t}$ denotes time in weeks. The post-ChatGPT indicator, $Post_{t}$, is set to 1 for weeks after the release of ChatGPT and 0 otherwise. The variable, $TimeAfter_{t}$, represents the number of weeks elapsed since the release of ChatGPT and is set to 0 during the pre-ChatGPT period.

In our study, the consumer group serves as the reference group. For consumers, the coefficients $\beta_0$, $\beta_1$, $\beta_2$, and $\beta_3$ represent, respectively, the estimated outcome level at the beginning of the observation window, the estimated weekly pre-ChatGPT trend, the immediate level change at ChatGPT’s release relative to the level projected by the pre-event trend, and the change in the weekly slope after the release. The group-specific coefficients capture differences relative to the reference group. For each non-reference group, $\gamma_{g}$ represents its baseline-level difference from consumers, $\phi_{g}$ represents its difference in the weekly pre-event trend, $\delta_{g}$ represents its difference in the immediate level change, and $\theta_{g}$ represents its difference in the post-event slope change. Thus, the model allows each user group to have a distinct baseline level and pre-ChatGPT trajectory, as well as different level and trend changes following the release of ChatGPT. Coefficients on the interaction terms indicate whether a given group's change differs significantly from that of the consumer reference group, rather than whether the change within that group differs significantly from zero. A group's own level or slope change is obtained by summing the corresponding reference-group coefficient and the group's interaction coefficient. 

%Standard errors and significance tests for these combined estimates are calculated using the model's full variance, thereby accounting for the covariance between the two coefficients.

\subsection{Similarity Between User Answers and AI-Generated Answers}
\label{user_ai_sim}
We examine whether changes in knowledge-production behavior are consistent with increased generative AI assistance. Specifically, we measure the semantic similarity between answers posted by Stack Overflow users and answers generated by ChatGPT for the same questions. We randomly selected 54,575 questions posted during our observation period and used OpenAI’s GPT-3.5 model to generate answers to these questions. We then used OpenAI's text-embedding-3-small model to measure the average semantic similarity between the answers posted by users and the AI-generated answers \cite{lyu2025wikipedia}. Appendix \ref{exp_user_ai_sim} presents examples of answers posted by users that are highly similar to AI-generated answers. For robustness, we also measured the cosine similarity between user answers and AI-generated answers for the same questions using the Sentence Transformers model \cite{reimers-gurevych-2019-sentence}, all-MiniLM-L6-v2. We found consistent results across the two models (see Appendix \ref{full_version}). For each week, we computed the average similarity score between AI-generated answers and user answers across user groups. We compare changes in this similarity score over time and across user groups. It should be noted that greater similarity is interpreted as evidence consistent with AI assistance, rather than as direct proof that a particular answer was generated using ChatGPT.

\subsection{Question Novelty and Complexity}
\label{ques_nov}
We collected the tags attached to questions in our dataset to measure question novelty across user groups based on the use of novel and rising tags. First, we identified 6,503 novel tags, defined as tags that appeared for the first time after the release of ChatGPT. To identify rising tags, we calculated the percentage growth in the frequency of each tag that had appeared before ChatGPT’s release. Specifically, we subtracted the tag’s pre-release frequency from its post-release frequency and divided the difference by its pre-release frequency. We classified the 5,086 tags in the top 10\% of this distribution as rising tags. The ten tags with the highest growth rates were large-language-model, apache-iotdb, iotdb, next.js13, .env, spring-boot-3, php-8.2, tanstack, micrometer-tracing, and sqlite3-python. Finally, for each user group, we calculated the proportions of questions containing novel and rising tags and tested whether these proportions differed across groups.

To measure question complexity, we trained a model to classify the difficulty of each question’s programming content, following prior research \cite{helic2026stack}. We used LeetCode problems labeled as Easy, Medium, or Hard as training examples and represented each problem using its title, description, and solution text. We converted these inputs into CodeT5 embeddings and used them to train an XGBoost classifier. We then applied the same embedding procedure to the textual and code content of Stack Overflow questions and used the trained classifier to assign each question a predicted complexity category.

\section{Results}

This section focuses on the comparison between consumers and producers. The complete estimates for all four user groups, including hybrid and low-activity users, are presented in Appendix \ref{full_version}.

\subsection{Consumers and Producers Changed the Volume and Distribution of Knowledge Consumption and Production Following the Release of ChatGPT}
\label{quantity_res}

Figure 1 reveals a divergence in the participation patterns of consumers and producers following the release of ChatGPT. Consumers reduced their question-asking activity but increased their answer production. Producers showed a contrasting pattern; their question activity remained stable, while their answer production declined substantially.

\begin{figure*}
  \centering
\includegraphics[width=\linewidth]{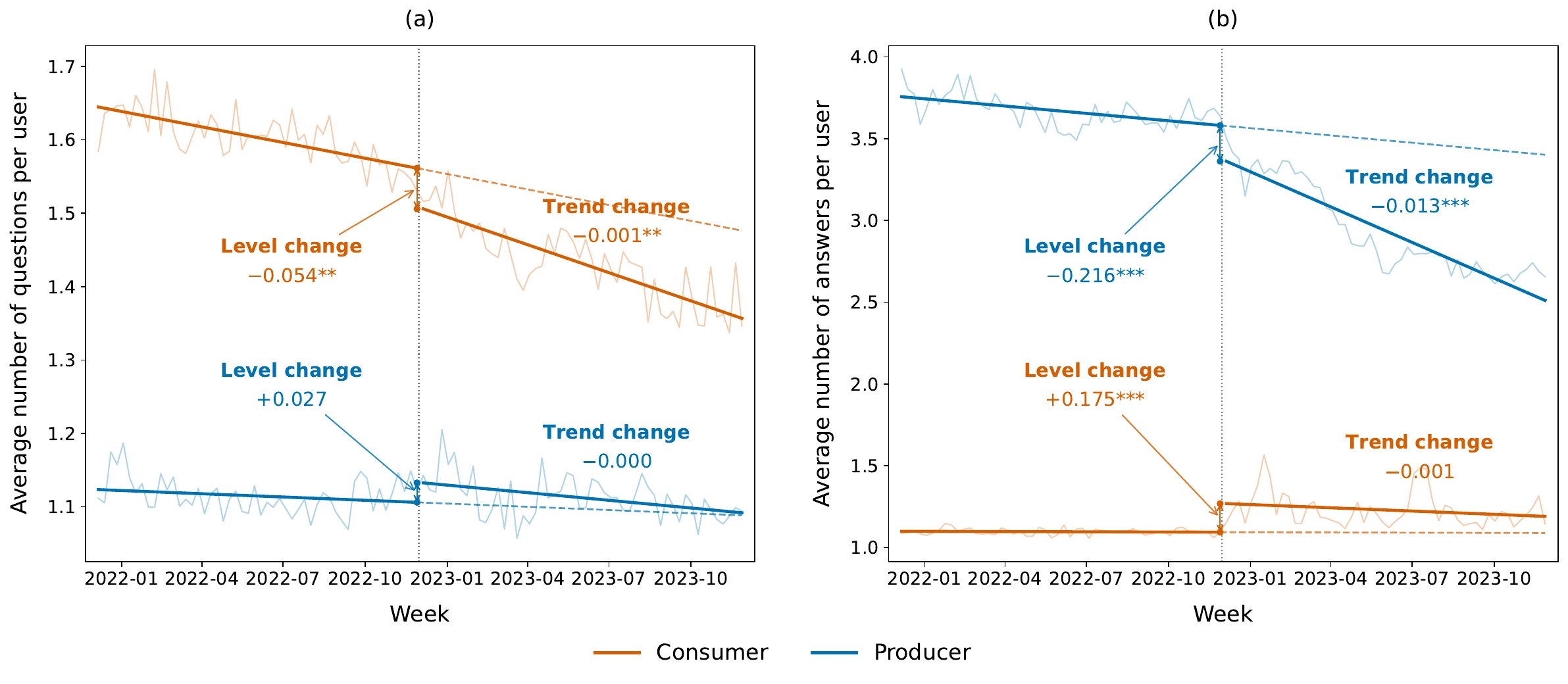}
  \caption{Panel (a) shows weekly trends in the average number of questions posted per user, and Panel (b) shows weekly trends in the average number of answers posted per user, separately for consumers and producers. The two subplots use different y-axis scales. The dotted vertical line marks the introduction of ChatGPT. Solid lines are fitted ITS trends; dashed lines extrapolate each group's pre-event trend forward as a counterfactual of what would have occurred absent the event. For each group, the annotated level change reports the immediate post-event shift relative to the counterfactual, while the annotated trend change reports the change in the weekly slope after the event relative to the pre-event slope. Estimates are tested against zero using HAC-robust standard errors.}
   \Description{Orange represents consumers and blue represents producers. Before ChatGPT, consumer questions declined gradually from about 1.64 to 1.56 per user, while producer questions remained near 1.1. After introduction, consumer questions show a downward level change of 0.054 and a further weekly trend change of −0.001, both marked statistically significant. Producer questions show a small upward level change of 0.027 and essentially no trend change; neither estimate has significance stars. For answers, producers began near 3.75 per user and were already declining slightly. They show a significant downward level change of 0.216 followed by a significant trend change of −0.013 per week, producing an increasingly large gap below the counterfactual and reaching a fitted value near 2.5 by late 2023. Consumer answers were approximately flat near 1.1 before ChatGPT, show a significant upward level change of 0.175, and then decline only slightly relative to the new level; the −0.001 trend change has no significance stars. Thus, the largest sustained post-event change is the decline in producer answers.}
\end{figure*}

% After the launch of ChatGPT, consumers posted significantly fewer questions and significantly more answers per user; producers posted significantly fewer answers, with the decline in answer volume accelerating further over time.

Figure 1(a) presents the results for question posting. At the beginning of the pre-ChatGPT period, consumers posted an estimated average of 1.64 questions per user, while producers posted significantly fewer questions. At the release of ChatGPT, consumers exhibited an immediate decline of questions per user ($\textit{b}=-0.054$, $\textit{p}<0.01$). Producers exhibited a small increase in the number of questions per user, estimated by summing the consumer-group coefficient ($b=-0.054$) and the producer-specific coefficient ($b=0.081$); however, this level change was not statistically significant. The groups also differed in how their trajectories changed after the release. Consumers experienced an additional downward change in their question-posting trend ($\textit{b}=-0.001$, $\textit{p}<0.01$). The corresponding trend change among producers was close to zero and not statistically significant. 

Figure 1(b) shows the contrasting pattern for answer production. At the beginning of the pre-ChatGPT period, consumers posted an estimated average of 1.10 answers per user. At the same point in time, producers posted an estimated 2.66 more answers per user than consumers ($\textit{b}=2.657$, $\textit{p}<0.001$), corresponding to approximately 3.76 answers per producer. At the release of ChatGPT, consumers' answer production increased immediately ($\textit{b}=0.175$, $\textit{p}<0.001$), whereas producers experienced an immediate decline, estimated by summing the consumer-group coefficient ($b=0.175$) and the producer-specific coefficient ($b=-0.391$). In addition, producers experienced an additional downward trend change.

To understand which consumers contributed to the change in answer volume, we divided consumers who answered during at least one period into three mutually exclusive groups. We excluded 3,405 consumers who answered in neither the pre- nor post-ChatGPT period. The three groups were continuing answerers, who answered both before and after ChatGPT’s release ($\textit{n}=$ 2,802); discontinued answerers, who answered only before the release ($\textit{n}=$ 4,506); and newly active answerers, who answered only after the release ($\textit{n}=$ 611). The average answer volume of continuing answerers remained nearly unchanged, decreasing slightly from 2.77 answers per user before the release of ChatGPT to 2.75 afterward. By contrast, many low-volume discontinued answerers, who averaged 1.83 answers each before the release, did not answer during the post-release period. Consequently, the more active continuing answerers accounted for 82\% of post-ChatGPT answerers. In addition, newly active answerers averaged 2.36 answers per user after the release, compared with 1.83 answers among discontinued answerers before the release. The increase in average answer volume among consumers therefore appears to reflect both the departure of relatively low-volume answerers and the participation of previously non-answering consumers with relatively high post-ChatGPT answer volumes.

\subsection{The Nature and Outcomes of Participation Changed Across User Groups}
\label{quality_res}

\subsubsection{Consumers produce more AI-like answers}

\begin{figure*}
  \centering
\includegraphics[width=0.5\linewidth]{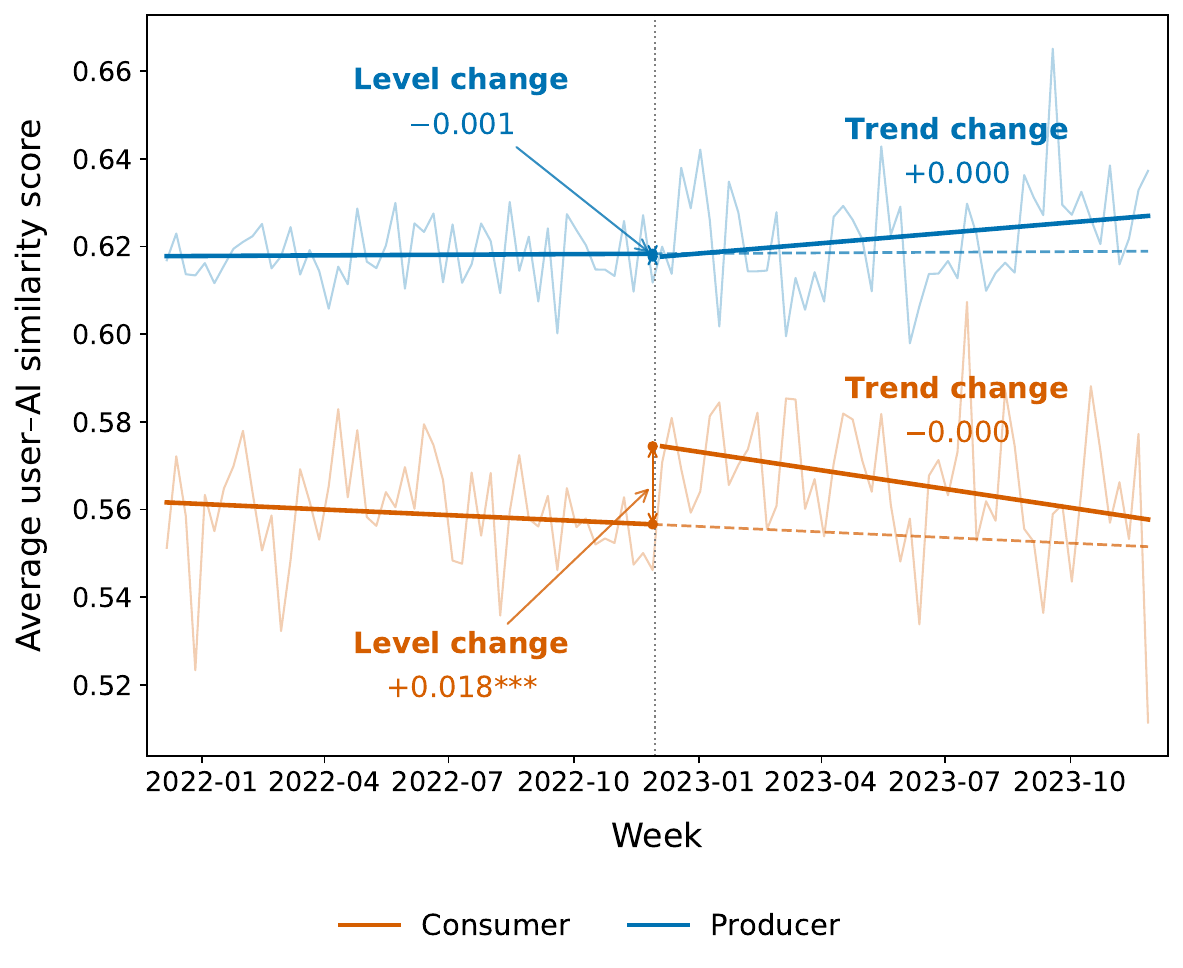}
  \caption{Weekly trends in the average semantic similarity between user answers and AI-generated answers, measured with OpenAI's text-embedding-3-small model, separately for consumers and producers. The dotted vertical line marks the introduction of ChatGPT. Solid lines are fitted ITS trends; dashed lines extrapolate each group's pre-event trend forward as a counterfactual of what would have occurred absent the event. For each group, the annotated level change reports the immediate post-event shift relative to the counterfactual, while the annotated trend change reports the change in the weekly slope after the event relative to the pre-event slope. Estimates are tested against zero using HAC-robust standard errors.}
  \Description{Orange represents consumers and blue represents producers. Producers have consistently higher similarity scores than consumers throughout the study period. Before ChatGPT, producer scores remain nearly flat around 0.62, while consumer scores decline slightly from about 0.562 to 0.557. At the event, consumers show a statistically significant level increase of 0.018, raising the fitted score to about 0.575. Their post-event fitted trend declines gradually, narrowing the initial increase relative to the counterfactual; the rounded trend change of −0.000 is not statistically significant. Producers show a negligible level change of −0.001 and a small positive post-event trend change that rounds to +0.000; neither is statistically significant. Their fitted score rises gradually to approximately 0.627 by late 2023.}
\end{figure*}

According to Figure 2, the average user–AI similarity score for consumers’ answers increased significantly at the launch of ChatGPT ($b=0.018$, $p<0.001$). Compared with the pre-GPT period, consumers posted answers that were more similar to AI-generated answers. In contrast, producer's similarity score barely changed after the launch. One possible explanation is that experienced answerers may be more likely to follow the community rule prohibiting the use of AI \cite{ai_ban_policy}. As a robustness check, we recomputed the semantic similarity between user and AI-generated answers using an alternative model based on Sentence Transformers and obtained results consistent with our main findings (see Appendix \ref{full_version}).

\subsubsection{The share of producers’ answers accepted by question askers increased, while the corresponding share for consumers declined}

Figure 3 reveals substantial heterogeneity in the share of answers accepted by question askers across user groups. For each user group and week, we calculate the share of answers accepted by dividing the total number of answers from that group that were accepted by the total number of answers provided by the group. An answer is counted as accepted when the asker of the question marks it as the accepted answer. Although consumers increased their answer production at the release of ChatGPT, the share of their answers accepted by question askers immediately decreased ($b=-0.020$, $p<0.05$). In contrast, producers experienced an immediate increase, estimated by summing the consumer-group coefficient ($b=-0.020$) and the producer-specific coefficient ($b=0.043$). The post-release trajectories further reinforced this divergence. Consumers experienced an additional negative change, indicating that their share of answers accepted continued to decline following the initial drop. Producers also exhibited a small negative trend change after the release, although its magnitude was substantially smaller than that of consumers. Together, these results suggest that consumers' increased answer production after ChatGPT did not translate into greater recognition. Instead, the acceptance gap between consumers and producers widened. 

\begin{figure*}
  \centering
\includegraphics[width=0.5\linewidth]{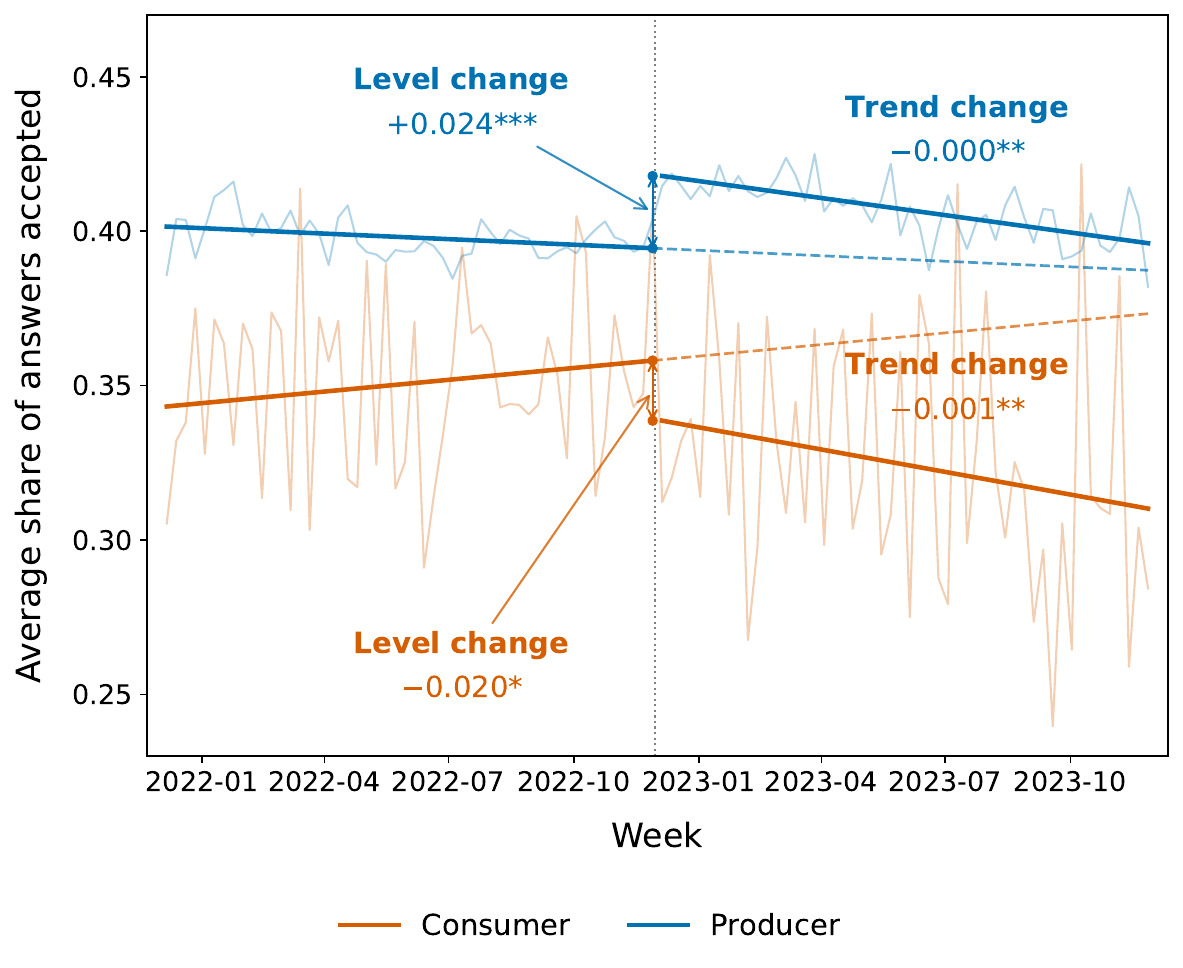}
  \caption{Weekly trends in the average share of answers accepted by question askers, separately for consumers and producers. The dotted vertical line marks the introduction of ChatGPT. Solid lines are fitted ITS trends; dashed lines extrapolate each group's pre-event trend forward as a counterfactual of what would have occurred absent the event. For each group, the annotated level change reports the immediate post-event shift relative to the counterfactual, while the annotated trend change reports the change in the weekly slope after the event relative to the pre-event slope. Estimates are tested against zero using HAC-robust standard errors.}
  \Description{Orange represents consumers and blue represents producers. Producers have a higher acceptance share throughout the study period. Before ChatGPT, the fitted consumer share rises from about 0.344 to 0.358, whereas the producer share declines slightly from about 0.402 to 0.395. At the event, consumers show a statistically significant level decrease of 0.020, followed by a significant trend change of −0.001 per week. Their fitted acceptance share subsequently falls to approximately 0.31 by late 2023, diverging from a counterfactual that continues upward to about 0.37. Producers show a statistically significant level increase of 0.024, to roughly 0.418, followed by a small but significant negative trend change that rounds to −0.000. Their fitted share gradually returns to approximately 0.396 by late 2023 but remains above the counterfactual.}
\end{figure*}

This divergence could arise from several mechanisms, including differences in the suitability or quality of the answers provided and changes in the types of questions that consumers chose to answer. For example, because consumers provided more AI-like answers than they did during the pre-ChatGPT period, one possible explanation for the lower share of their answers accepted is that their answers are more likely to contain inaccuracies \cite{ai_ban_policy, kabir2024stack}. We conducted follow-up analyses to investigate other potential mechanisms (see Appendix \ref{poten_mech}). Neither a shift toward more difficult or abandoned questions nor increased competition from other answers appears to account for the post-ChatGPT decline in the share of consumers’ answers accepted. Thus, the decline may instead reflect inaccuracies in consumers’ answers or a growing preference among question askers for answers from established producers.

\subsubsection{Producers remain better positioned to ask questions about novel and emerging topics}

Generative AI also widens the gap between novice and experienced users in terms of how they use online communities to either outsource basic tasks or master cutting-edge expertise. We examined the tags attached to questions to understand what types of questions users have posted since the launch of ChatGPT. Pairwise proportion \textit{z}-tests with Bonferroni correction indicated significant differences between all group pairs for both novel and rising tags (see Table \ref{tab5}). Producers exhibited the highest proportion of questions containing both novel and rising tags. In other words, following the release of ChatGPT, producers were more likely than other groups to ask questions involving novel, rising topics (e.g., large language models). 

\begin{table*}
\centering
\caption{Proportions of questions containing at least one novel tag and of questions containing at least one rising tag, by user group. Pairwise proportion \textit{z}-tests with Bonferroni correction compare each group with the producer group. Differences are percentage-point differences relative to the producer group. $^{*}p<0.05$; $^{**}p<0.01$; $^{***}p<0.001$.}
\label{tab5}
\renewcommand{\arraystretch}{1.1}
\begin{tabular}{l ll ll}
\toprule
 & \multicolumn{2}{c}{Novel tag} & \multicolumn{2}{c}{Rising tag} \\
\cmidrule(lr){2-3} \cmidrule(lr){4-5}
Group & Share (\%) & Difference from Producer & Share (\%) & Difference from Producer \\
\midrule
Consumer     & 1.90 & $-1.45^{***}$ & 6.23 & $-2.62^{***}$ \\
Hybrid       & 2.89 & $-0.45^{**}$  & 8.15 & $-0.69^{**}$ \\
Low-activity & 2.40 & $-0.94^{***}$ & 7.41 & $-1.44^{***}$ \\
Producer     & 3.34 &            & 8.85 &            \\
\bottomrule
\end{tabular}
\Description{Producers have the highest proportion of questions containing novel or rising tags. For novel tags, the shares are 1.90\% for consumers, 2.89\% for hybrids, 2.40\% for low-activity users, and 3.34\% for producers. Relative to producers, the differences are −1.45 percentage points for consumers (p < 0.001), −0.45 for hybrids (p < 0.01), and −0.94 for low-activity users (p < 0.001). For rising tags, the shares are 6.23\% for consumers, 8.15\% for hybrids, 7.41\% for low-activity users, and 8.85\% for producers. Relative to producers, the differences are −2.62 percentage points for consumers (p < 0.001), −0.69 for hybrids (p < 0.01), and −1.44 for low-activity users (p < 0.001). Thus, every comparison group has a significantly smaller share than producers for both tag categories.}
\end{table*}

\section{Discussion}
\label{discussion}
This research offers several implications that advance our understanding of the human behaviors, motivations, and interactions underlying online knowledge seeking and production. We also derive design implications for supporting different user groups equitably and offer recommendations for knowledge-sharing platforms responding to AI-driven changes in users’ participation patterns and outcomes.

\subsection{Theoretical Contributions}
\label{theory}

\subsubsection{Generative AI complicates the characterization of consumers as free riders}

Our findings suggest that generative AI changes how knowledge consumers and producers participate within online knowledge ecosystems. The expansion of consumers’ answer production following the release of ChatGPT challenges the conventional characterization of consumers as free riders who benefit from collective knowledge without reciprocating. Previous research using pre-ChatGPT Stack Overflow data documents a negative relationship between knowledge seeking and contribution among active, consistent users, attributing this pattern to free riding or a limited motivation to reciprocate \cite{mustafa2023motivates}. Our findings point to an alternative explanation; some consumers may have been willing to contribute but lacked the knowledge, confidence, or resources needed to formulate answers. By reducing these barriers, generative AI may enable such users to act on previously constrained motivations to reciprocate. 

This interpretation qualifies theories of free riding and the tragedy of the digital commons \cite{li2011factors}. Low observed contribution does not necessarily indicate an absence of motivation; it may instead reflect users’ limited capacity to participate. By lowering barriers to contribution, generative AI may allow previously latent willingness to reciprocate to become observable. This finding also extends research on reciprocity in knowledge-sharing communities \cite{wu2013you} and the social contagion of prosocial behavior \cite{tsvetkova2014social}. Whereas prior work has emphasized social mechanisms that encourage reciprocity, our findings suggest that technological assistance can also facilitate contributions motivated by reciprocity and by altruistic or strategic goals \cite{lou2013contributing, lai2014knowledge, luo2021effect, licorish2026you}. Theories of online knowledge sharing should therefore avoid treating consumers and producers as fixed user types. Instead, these roles should be understood as participation patterns that depend partly on the technological resources shaping users’ capacity and incentives to contribute.

\subsubsection{Generative AI expands participation but does not yield equal benefits}

We also found that consumers' expanded production does not yield greater recognition. Although consumers increased their answer production after the release of ChatGPT, this increase was not accompanied by a proportional rise in accepted answers. This pattern suggests that reducing the cost of producing content is insufficient to equalize epistemic standing within a knowledge-sharing community. The lower share of their answers accepted may be consistent with concerns about the accuracy and verification of AI-assisted content. Such concerns were central to Stack Overflow’s policy banning AI-generated content, adopted on December 5, 2022 \cite{ai_ban_policy}. This interpretation also aligns with a previous finding that 52\% of ChatGPT-generated answers to 517 Stack Overflow questions contained incorrect information \cite{kabir2024stack}. Experimental evidence further suggests that overreliance on AI may impair users’ ability to understand, read, and debug code \cite{shen2026ai}. Recognition may continue to depend on accumulated expertise, reputation, and the ability to tailor answers to the platform’s norms \cite{movshovitz2013analysis}. More broadly, the finding distinguishes contribution quantity from recognized contribution quality; AI may help less-experienced answerers produce more answers, but additional social and technical mechanisms may be needed to help them produce answers that the community recognizes as valuable. 

Furthermore, we found that producers retain an advantage through greater engagement with novel and emerging topics. One interpretation is that generative AI accelerates experienced users’ routine work, allowing them to progress more quickly toward advanced problems that existing knowledge cannot readily resolve. This finding also highlights an important asymmetry in knowledge consumption. Whereas generative AI may reduce the barriers to generating answers, identifying valuable frontier questions may continue to require accumulated expertise, domain-specific judgment, and an understanding of the boundaries of existing knowledge. 

\subsection{Design Implications}
\label{design}
Our findings invite a broader reconsideration of what online knowledge communities should provide when generative AI can increasingly satisfy users' routine information needs. Rather than simply preserving existing patterns of asking and answering, future platforms may need to redefine the division of epistemic labor among AI systems, information seekers, and human experts. Because technological affordances and artifacts can promote active participation \cite{khansa2015understanding}, we identify four design opportunities for knowledge-sharing communities.

First, platforms should evolve from repositories of existing knowledge into infrastructures that support the creation and validation of frontier knowledge. As generative AI increasingly handles established, frequently asked, and well-documented questions, the comparative value of online knowledge communities may increasingly lie in supporting knowledge that is not yet readily available to AI \cite{liu2023chatgpt}. Our finding that prior producers increasingly ask questions about novel and emerging topics in the wake of ChatGPT points toward such a shift. Future platforms could identify and elevate questions concerning new technologies, unresolved problems, rapidly changing information, or topics for which existing knowledge remains sparse. They might also create dedicated spaces for collective sensemaking around emerging topics, where the goal is not necessarily to immediately produce a canonical answer but to collectively explore and stabilize new knowledge. More broadly, knowledge-sharing platforms might shift from optimizing primarily for efficient retrieval of what is already known toward facilitating the discovery, discussion, validation, and accumulation of what is not yet known.

Second, platforms should help users move beyond simply delivering AI-generated answers and instead use AI to develop their capacity as human knowledge producers. Our finding that consumers provide more answers after the release of ChatGPT suggests that generative AI may lower the barrier to knowledge production. Yet their declining share of answers accepted indicates that increased participation does not necessarily translate into valued contribution. Rather than simply helping users generate answers that can be copied and pasted into a community, platforms could use AI to scaffold users' development as knowledge contributors. For example, AI-assisted answering interfaces could help users locate supporting documentation, verify code, identify unsupported claims, compare proposed answers with existing knowledge, or incorporate firsthand evidence and experience. Given that personalized recommendation better supports knowledge sharing \cite{zhang2019understanding}, platforms could also recommend questions that are closely relevant but slightly beyond users' current expertise and provide AI support as they attempt to answer them. The design goal would therefore shift from using AI to produce answers for users toward using AI to help users become capable knowledge producers, potentially turning participation itself into a pathway for developing expertise.

Third, platform designers should replace competition between AI and human communities with an intelligent division of epistemic labor. Future information systems could help users determine not only how to answer a question, but where that question should be answered. Stable, well-documented, and repeatedly answered questions may be efficiently handled by generative AI or existing community archives, whereas novel, rapidly evolving, contested, experience-dependent, or poorly documented questions may benefit more from human communities. Platforms could therefore develop routing mechanisms that direct users among AI assistance, existing community knowledge, and new community discussion based on the characteristics of their information needs. AI systems could similarly recognize when their knowledge is insufficient or outdated and direct users to relevant human communities. This design principle suggests a complementary human–AI–community knowledge pipeline; AI helps users resolve known components and articulate unresolved ones, communities generate and validate emerging knowledge, and this newly established knowledge can subsequently support future users and AI systems.

Finally, practitioners should redesign credit and incentive systems to recognize the distinctive epistemic value that human contributors provide. As AI lowers the cost of producing plausible answers, platforms may also need to reconsider which types of contributions deserve recognition. Rather than rewarding answers simply for being human-written or differing from AI-generated text, communities could recognize what contributors add beyond readily available AI-generated knowledge, such as firsthand experience, empirical verification, emerging knowledge, counterexamples, corrections, or the discovery of novel problems. Interfaces might even indicate what AI can already provide and prompt contributors to identify what they can add. Recognition systems could similarly reward verification, correction, synthesis, and problem discovery rather than concentrating credit solely on the final accepted answer. 

%This approach may be particularly important given our finding that consumers contributed more after the release of ChatGPT but received less recognition, suggesting that broader participation alone does not ensure a more equitable distribution of recognition.

\subsection{Limitations and Future Work}

This study is not without limitations. First, we could not incorporate certain quantitative metrics, such as reputation scores and upvotes, into this study due to data limitations. Since the dump dataset contains aggregated values that are cumulative up to the dump creation date, we could not analyze how reputation scores and upvote counts changed over time. Hence, we chose acceptance as a measure of recognition instead of upvote. It should be noted that acceptance is determined by a single asker, not the broader community. We decided to adopt a clearer signal (i.e., whether an answer was accepted by the asker) instead of an unstable one. 

Another limitation is that it is extremely difficult to determine whether an answer author used ChatGPT when composing a post. We attempted to capture potential AI use by measuring the similarity between answers generated by GPT-3.5 and those posted by users. Although the patterns of textual overlap reported in the Appendix \ref{exp_user_ai_sim} are suggestive, this measure is an imperfect proxy for actual AI use. For example, independently written answers may resemble AI-generated responses because they use common technical language, whereas heavily edited AI-assisted answers may exhibit little textual similarity. Prior studies have proposed several methods for detecting AI-generated content \cite{brooks2024rise, mitchell2023detectgpt}, but such detectors also remain vulnerable to false positives, false negatives, model drift, and strategic editing. Consequently, our results should not be interpreted as direct estimates of individual-level ChatGPT adoption. This measurement limitation also constrains causal interpretation. Because we cannot reliably distinguish treated users who adopted generative AI from control users who did not, our analyses identify behavioral changes associated with the launch of ChatGPT rather than the causal effect of AI use itself. Moreover, the platform’s AI-content ban was introduced during the same period and may have independently affected participation, moderation, and answer-posting behavior \cite{borwankar2026unraveling}. Future research could exploit settings with more precise measures of AI adoption, staggered policy implementation, or exogenous variation in access to generative AI to better isolate causal effects.

Next, our study population is restricted to users with sufficient activity records during the pre-ChatGPT period. This design enables us to compare the same users’ behavior over time, but it excludes lurkers and users who newly joined after the release of ChatGPT. The resulting sample is therefore more representative of established and relatively active users than of the platform’s entire user population. The findings may consequently have limited generalizability to newer or less active users and to other online knowledge communities. In addition, we classify users primarily according to their prior activity patterns. Although this approach captures meaningful differences in platform roles, users may also exhibit heterogeneous responses based on gender, geography, tenure, and other characteristics \cite{bachschi2020asking}. Future work could examine intersections between behavioral roles and demographic or contextual characteristics, provided that such analyses can be conducted with appropriate attention to privacy and statistical power.

Finally, our observational analysis documents role-specific behavioral patterns but cannot fully explain the motivations underlying them. For example, a decline in answer posting could reflect dissatisfaction with platform policies or reduced perceived value from contributing. Future research could combine behavioral data with interviews, surveys, or qualitative analyses of user discussions to examine how contributors perceive generative AI and how those perceptions shape their decisions to ask, answer, or leave the platform. Such mixed-methods evidence would help distinguish among competing explanations and provide a more complete account of how generative AI is reshaping participation in online knowledge communities.

\section{Conclusion}

Online knowledge-sharing platforms are interdependent knowledge infrastructures whose sustainability depends on complementary roles in seeking and producing information. Because generative AI can substitute for community knowledge while also lowering barriers to producing it, its effect is theoretically ambiguous and may depend on a user’s prior role. Although prior research has primarily examined aggregate changes in traffic and contribution volume, these trends can conceal offsetting shifts between user groups. We use a role-based framework to examine how the release of ChatGPT was associated with changes in knowledge seeking and production among consumers, producers, hybrid users, and low-activity users on Stack Overflow, with a particular focus on differences between consumers and producers. We find a reversal of traditional participation patterns following the release of ChatGPT. Consumers asked fewer questions but provided more answers, whereas producers provided fewer answers. Consumers’ increased answer production was accompanied by greater similarity to AI-generated responses, consistent with generative AI lowering barriers to contribution. However, these role shifts did not produce comparable outcomes across groups. Consumers' answers were accepted less often, whereas producers received greater recognition for their answers and remained more likely to ask questions about emerging and novel topics. These findings suggest that generative AI may enable users to participate beyond their established roles without necessarily narrowing gaps in the benefits associated with such participation.

%\begin{acks}
%To Robert, for the bagels and explaining CMYK and color spaces.
%\end{acks}

%\section*{Ethics and Privacy Statement}

%This study uses publicly available Stack Overflow data obtained from the Internet Archive, involves no direct interaction with human participants, and reports results only in aggregate. The dataset contains no sensitive personal information. Because measures of AI-like language cannot establish individual AI use, our findings should not be used to infer or police users' use of AI or to stigmatize particular user groups.

\bibliographystyle{ACM-Reference-Format}
\bibliography{sample-base}

\appendix

\section{Sensitivity Analysis}
\label{sen_anal}
To assess the sensitivity of our findings to the thresholds used for identifying active users and assigning behavioral roles, we examined alternative minimum-activity and role-assignment criteria. Our primary specification requires at least 10 actions for a user to be classified as active and assigns a consumer or producer role when at least 80\% of the user’s actions correspond to that activity type. These thresholds prioritize reliable classification by ensuring that roles are based on sufficient activity and a clearly dominant behavioral pattern.

% A minimum-activity threshold of 20 assigns 98.3\% of users to the low-activity group, leaving comparatively small samples in the consumer, hybrid, and producer roles. This imbalance motivates our choice of smaller minimum activity thresholds.

\begin{table*}
\centering
\caption{Sensitivity of role distributions to minimum-activity and role-assignment thresholds. Each cell reports the number of users assigned to a role, with the percentage of the full sample in parentheses.}
\label{tab3}
\renewcommand{\arraystretch}{1.1}
\begin{tabular}{lllll}
\toprule
Minimum activity & Role & 60\% & 70\% & 80\% \\
\midrule
5 & Consumer & 54,985 (5.9\%) & 45,424 (4.9\%) & 39,408 (4.3\%) \\
 & Hybrid & 7,761 (0.8\%) & 23,501 (2.5\%) & 33,286 (3.6\%)\\
 & Low-activity & 811,977 (87.6\%) & 811,977 (87.6\%) & 811,977 (87.6\%)\\
  & Producer & 52,016 (5.6\%) & 45,837 (4.9\%) & 42,068 (4.5\%)\\
\midrule
10 & Consumer & 16,262 (1.8\%) & 14,064 (1.5\%) & 11,324 (1.2\%) \\
 & Hybrid & 2,961 (0.3\%) & 6,750 (0.7\%) & 11,383 (1.2\%)\\
 & Low-activity & 883,382 (95.3\%)  & 883,382 (95.3\%)  & 883,382 (95.3\%) \\
  & Producer & 24,134 (2.6\%) & 22,543 (2.4\%) & 20,650 (2.2\%)\\
\midrule
20 & Consumer & 4,261 (0.5\%) & 3,724 (0.4\%)& 2,990 (0.3\%) \\
 & Hybrid & 801 (0.1\%) & 1,779 (0.2\%)& 3,097 (0.3\%)\\
 & Low-activity & 910,715 (98.3\%) & 910,715 (98.3\%) & 910,715 (98.3\%)\\
  & Producer & 10,962 (1.2\%) & 10,521 (1.1\%) & 9,937 (1.1\%)\\
\bottomrule
\end{tabular}
\Description{Raising the minimum-activity requirement shifts most users into the low-activity group: 87.6\% with a 5-action minimum, 95.3\% with 10 actions, and 98.3\% with 20 actions. These percentages do not vary with the role-assignment threshold. At each activity minimum, increasing the role threshold from 60\% to 80\% reduces the consumer and producer groups while enlarging the hybrid group. Under the primary specification of at least 10 actions and an 80\% role threshold, 11,324 users (1.2\%) are consumers, 11,383 (1.2\%) are hybrids, 883,382 (95.3\%) are low-activity, and 20,650 (2.2\%) are producers. Relaxing both criteria to 5 actions and 60\% yields the largest consumer and producer groups. The most restrictive combination, 20 actions and 80\%, leaves only 2,990 consumers (0.3\%), 3,097 hybrids (0.3\%), and 9,937 producers (1.1\%), with 910,715 users classified as low-activity. Across all threshold combinations, producers generally outnumber consumers, except under the least restrictive 5-action, 60\% specification.}
\end{table*}

The choice of thresholds involves a trade-off between sample size and classification precision. Lowering the minimum-activity threshold includes more users in the consumer, producer, and hybrid groups, but classification results based on fewer observed actions may be less stable. Similarly, lowering the role-assignment threshold increases the number of users classified as consumers or producers, but it weakens the distinction between these roles because users with more mixed behavioral patterns may be assigned to a dominant role rather than classified as hybrid.

We also considered a minimum-activity threshold of 20 actions. As shown in Table \ref{tab3}, however, this threshold assigns 910,715 users (98.3\% of the sample) to the low-activity group, regardless of the role-assignment proportion. Consequently, the consumer, producer, and hybrid groups become extremely small, reducing statistical power and limiting reliable comparisons across roles. Because these comparisons are central to our analysis, thresholds above 10 actions provide little analytical benefit.

We therefore retain the specification requiring at least 10 actions and an 80\% role-assignment threshold as our primary definition and assess robustness using the other three combinations of the two parameter values: 5 actions and 80\%, 10 actions and 60\%, and 5 actions and 60\%. This design allows us to examine the effects of relaxing the minimum-activity threshold and the role-assignment threshold both separately and jointly. Across all four specifications, however, the results remain substantively consistent, indicating that our conclusions are not sensitive to the particular thresholds used to define active users and assign behavioral roles.

% Lower thresholds increase the sizes of the consumer, producer, and hybrid groups but may reduce classification precision by relying on fewer observed actions or allowing users with less clearly dominant behavioral patterns to be assigned to a role. 

\section{Examples of Stack Overflow Answers with High User–AI Similarity}
\label{exp_user_ai_sim}
This section presents the five Stack Overflow answers that have the highest semantic similarity to AI-generated answers.

\subsection{Example 1: User–AI Similarity = 0.937}

\begin{answerbox}{User Answer}
To make sure that \texttt{\$sentTo} is an empty collection when both
conditions are false...

\begin{Verbatim}[breaklines=true,breakanywhere=true,fontsize=\small]
$admins = collect(new User);
$owner = collect(new User);

if (config('personal.mailing_technician')) {
    $owner = $event->ticket->ownedBy;
}

if (config('personal.mailing_admin')) {
    $admins = User::query()->role('admin')->get();
}

$sentTo = collect(); // Initialize an empty collection

if ($admins->isNotEmpty() || $owner->isNotEmpty()) {
    $sentTo = $admins->push($owner);
}

if ($sentTo->isNotEmpty()) {
    // Do something
}
\end{Verbatim}
\end{answerbox}

\begin{answerbox}{AI-Generated Answer}
To ensure that \texttt{\$sentTo} is an empty collection when both
conditions are are false, you can modify the code as follows:

\begin{Verbatim}[breaklines=true,breakanywhere=true,fontsize=\small]
$admins = collect(new User);
$owner = collect(new User);

if (config('personal.mailing_technician')) {
    $owner = $event->ticket->ownedBy;
}

if (config('personal.mailing_admin')) {
    $admins = User::query()->role('admin')->get();
}

$sentTo = collect();

if ($owner->isNotEmpty() || $admins->isNotEmpty()) {
    $sentTo = $admins->push($owner);
}

if ($sentTo->isNotEmpty()) {
    // Do something if the collection is not empty
    .......
}
\end{Verbatim}

In the code above, we first initialize \texttt{\$sentTo} as an empty
collection using \texttt{collect()}. Then, we check if either
\texttt{\$owner} or \texttt{\$admins} contains any items. If either of
them has items, we populate \texttt{\$sentTo} using \texttt{push}.
Otherwise, \texttt{\$sentTo} remains an empty collection. This way,
\texttt{\$sentTo} will be empty if both conditions are false.
\end{answerbox}

\subsection{Example 2: User–AI Similarity = 0.918}

\begin{answerbox}{User Answer}
Use a filter to select only the rows where the meat type is
``chicken'', and then use the ``max'' function to find the row with
the highest price.

\begin{Verbatim}[breaklines=true,breakanywhere=true,fontsize=\small]
import pandas as pd

df = pd.DataFrame({
    'Meat type': ['beef', 'beef', 'chicken', 'chicken', 'pork', 'pork'],
    'brand': ['A', 'B', 'A', 'B', 'A', 'B'],
    'Price': [10, 12, 11, 14, 9, 16]
})

chicken_df = df[df['Meat type'] == 'chicken']

max_row = chicken_df.loc[chicken_df['Price'].idxmax()]

print(max_row)
\end{Verbatim}

\noindent\textit{Out:}

\begin{Verbatim}[breaklines=true,breakanywhere=true,fontsize=\small]
Meat type    chicken
brand              B
Price             14
Name: 3, dtype: object
\end{Verbatim}
\end{answerbox}

\begin{answerbox}{AI-Generated Answer}
You can achieve this by filtering the dataframe to only include rows
where the ``Meat type'' is ``chicken'', and then finding the row with
the maximum price in that subset. Here is a code snippet to accomplish
this using pandas:

\begin{Verbatim}[breaklines=true,breakanywhere=true,fontsize=\small]
import pandas as pd

data = {
    'Meat type': ['beef', 'beef', 'chicken', 'chicken', 'pork', 'pork'],
    'brand': ['A', 'B', 'A', 'B', 'A', 'B'],
    'Price': [10, 12, 11, 14, 9, 16]
}

df = pd.DataFrame(data)

max_chicken_price_row = (
    df[df['Meat type'] == 'chicken']
    .nlargest(1, 'Price')
)

print(max_chicken_price_row)
\end{Verbatim}

This code will output:

\begin{Verbatim}[breaklines=true,breakanywhere=true,fontsize=\small]
  Meat type brand  Price
3   chicken     B     14
\end{Verbatim}

This result is the row with the highest price for the ``chicken'' meat
type.
\end{answerbox}

\subsection{Example 3: User–AI Similarity = 0.904}

\begin{answerbox}{User Answer}
In the provided code, a and b represent the elements present in the heap
(PriorityQueue). The lambda expression
\texttt{(a, b) -> map.get(b) - map.get(a)} is used as the comparator for
the PriorityQueue, which determines how elements are compared and ordered
in the heap.

In a PriorityQueue, the elements are sorted based on the result of the
comparator. The comparator is applied to the elements present in the queue
to determine their relative order. In this case, the lambda expression
compares the values of the keys in the map associated with elements a and b.

The lambda expression \texttt{(a, b) -> map.get(b) - map.get(a)} subtracts
the value associated with element a from the value associated with element
b. This means that the element with a higher value in the map will be
considered greater, and therefore, will have a higher priority in the
PriorityQueue. Consequently, the PriorityQueue will act as a max heap,
where the element with the highest value in the map will be at the root
(top) of the heap.

In other words, the lambda expression is used to define the priority or
ordering of elements in the PriorityQueue, and by subtracting the values
in reverse order (b - a), we create a max heap where the element with the
highest value in the map has the highest priority.
\end{answerbox}

\begin{answerbox}{AI-Generated Answer}
In the lambda expression
\texttt{(a, b) -> map.get(b) - map.get(a)}, \texttt{a} and \texttt{b}
are elements from the priority queue that are being compared to determine
their order in the heap.

In this context, the lambda expression is used as a comparator for the
Priority Queue. When elements are added to the priority queue, the lambda
expression is used to determine the priority of the elements based on the
values retrieved from a map.

The lambda expression compares the values associated with elements
\texttt{a} and \texttt{b} in the map. The subtraction
\texttt{map.get(b) - map.get(a)} decides the order of elements in the
priority queue. If the result is positive, \texttt{b} will have a higher
priority than \texttt{a}, making it a max heap. If negative, \texttt{a}
will have a higher priority, making it a min heap.

So, \texttt{a} and \texttt{b} are elements being compared, and their
priority in the heap is determined by the values associated with them in
the map.
\end{answerbox}

\subsection{Example 4: User–AI Similarity = 0.904}

\begin{answerbox}{User Answer}
The NC in \texttt{WM\_NCCREATE} stands for "non-client." The non-client
area of a window in a Windows application is the application icon, menu
bar, title bar, minimize/maximize/close buttons, and resizeable border.
I.e. the parts of the window that the system draws. The
\texttt{WM\_NCCREATE} window message is sent prior to the
\texttt{WM\_CREATE} message when a window is first created.
\end{answerbox}

\begin{answerbox}{AI-Generated Answer}
In the context of Windows operating system programming, the "NC" in
\texttt{WM\_NCCREATE} stands for "non-client." This refers to the
non-client area of a window, which includes elements such as the window
border, title bar, menu bar, etc. The \texttt{WM\_NCCREATE} message is
sent to the window procedure of a newly created window before the window
is created, allowing the application to perform initialization tasks
related to the non-client area of the window.
\end{answerbox}

\subsection{Example 5: User–AI Similarity = 0.897}

\begin{answerbox}{User Answer}
You can modify your loop slightly to avoid adding a comma after the last
number. You can achieve this by checking if you're at the last iteration
of the loop. Here's an updated version of your function:

\begin{Verbatim}[breaklines=true,breakanywhere=true,fontsize=\small]
def get_fibonacci_sequence(num: int) -> str:
        '''
        Function to return string with Fibonacci sequence

        >>> get_fibonacci_sequence(0)
        ''
        >>> get_fibonacci_sequence(1)
        '0'
        >>> get_fibonacci_sequence(9)
        '0,1,1,2,3,5,8,13,21'
        '''

        if num <= 0:
            return ''  # Return an empty string for non-positive input

        first_term = 0
        second_term = 1
        nth_term = 0
        result = ''

        for sequence in range(num):
            result += str(first_term)
            if sequence < num - 1:  # Check if it's not the last iteration
                result += ','
            nth_term = first_term + second_term
            first_term = second_term
            second_term = nth_term

        return result

    # Test cases
    print(get_fibonacci_sequence(0))  # Should print ''
    print(get_fibonacci_sequence(1))  # Should print '0'
    print(get_fibonacci_sequence(9))  # Should print '0,1,1,2,3,5,8,13,21'
\end{Verbatim}
\end{answerbox}

\begin{answerbox}{AI-Generated Answer}
You can modify the function slightly to achieve the desired behavior.
You can check if you are on the last iteration of the loop and avoid
appending a comma in that case. Here's the updated code:

\begin{Verbatim}[breaklines=true,breakanywhere=true,fontsize=\small]
def get_fibonacci_sequence(num: int) -> str:
    '''
    Function to return string with fibonacci sequence

    >>> get_fibonacci_sequence(0)
    ''
    >>> get_fibonacci_sequence(1)
    '0'
    >>> get_fibonacci_sequence(9)
    '0,1,1,2,3,5,8,13,21'
    '''

    first_term = 0
    second_term = 1
    nth_term = 0
    result = ''

    for sequence in range(num):
        result += str(first_term)
        if sequence < num - 1:
            result += ','
        nth_term = first_term + second_term
        first_term = second_term
        second_term = nth_term
    return result

print(get_fibonacci_sequence(9))
\end{Verbatim}

This code checks if the current iteration is not the last one (by
comparing \texttt{sequence} with \texttt{num - 1}), and only appends a
comma in that case. This way, you get the Fibonacci sequence with commas
between the numbers and no extra comma at the end.
\end{answerbox}

\section{Potential Mechanisms Behind the Decline in Acceptance of Consumers' Answers}
\label{poten_mech}
This section presents the results of follow-up analyses testing potential mechanisms behind the decline in the share of consumers' answers accepted after the launch of ChatGPT.

First, it is possible that consumers may have begun answering a different mix of questions after the release of ChatGPT. For example, they may have shifted toward more complex or abandoned questions. We first examined whether consumers increasingly targeted difficult questions after the release of ChatGPT. We tracked the weekly share of each group's answers that targeted a question flagged as hard by our classifier. According to Figure 4(a), consumers showed no significant immediate level change but exhibited a small positive change in their post-ChatGPT trend. Producers showed both a significant immediate level increase and a small positive trend change. Although both groups increasingly targeted difficult questions, the changes were not unique to consumers. We also examined whether the questions answered by each group became increasingly likely to remain without an accepted answer. Figure 4(b) shows that neither consumers nor producers experienced a significant post-ChatGPT level change or trend change in this outcome. These findings suggest that neither increased targeting of difficult questions nor an increased likelihood that answered questions remained unresolved explains the consumer-specific decline in the share of answers accepted.

\begin{figure*}
  \centering
\includegraphics[width=\linewidth]{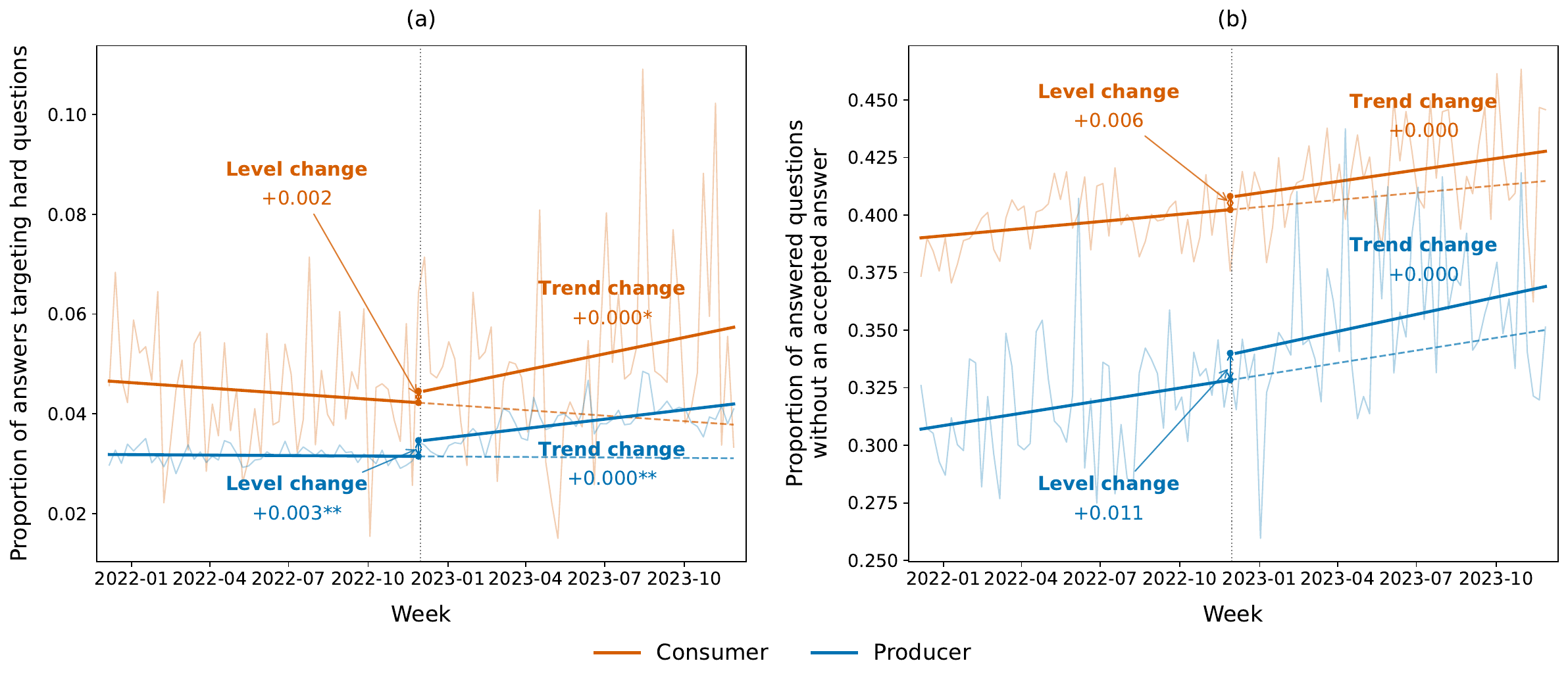}
  \caption{Panel (a) shows weekly trends in the average proportion of answers targeting hard questions, and Panel (b) shows weekly trends in the average proportion of answered questions wihtout an accepted answer, separately for consumers and producers. The dotted vertical line marks the introduction of ChatGPT. Solid lines are fitted ITS trends; dashed lines extrapolate each group's pre-event trend forward as a counterfactual of what would have occurred absent the event. For each group, the annotated level change reports the immediate post-event shift relative to the counterfactual, while the annotated trend change reports the change in the weekly slope after the event relative to the pre-event slope.}
   \Description{Orange represents consumers and blue represents producers. In panel (a), consumers have a higher fitted share of answers targeting hard questions throughout. Before ChatGPT, the consumer share declines from about 0.046 to 0.042, while the producer share remains near 0.032. At the event, consumers show a nonsignificant level increase of 0.002, whereas producers show a significant increase of 0.003. Both groups then experience small positive, statistically significant trend changes, although the displayed estimates round to +0.000. By late 2023, fitted shares reach approximately 0.058 for consumers and 0.042 for producers, both above their declining or nearly flat counterfactuals. In panel (b), consumers also have the higher proportion of answered questions without an accepted answer. Both groups trend upward before ChatGPT. The event is associated with level increases of 0.006 for consumers and 0.011 for producers, neither statistically significant. Positive trend changes for both groups round to +0.000 and are also not significant. Nevertheless, fitted post-event trajectories rise somewhat faster than the counterfactuals, reaching about 0.428 for consumers and 0.369 for producers by late 2023.}
\end{figure*}

The next possible scenario is that consumers increasingly compete with more existing answers already posted by the time they respond. Previous research has shown that faster answers are more likely to be accepted on Stack Overflow, regardless of their quality \cite{lu2020haste}. Thus, competing with more existing answers may mechanically reduce the likelihood that a consumer’s answer is selected. We measured this competition directly as the number of answers already present for a question at the moment each answer was posted. Contrary to this explanation, the average number of competing answers dropped immediately after the release of ChatGPT for both consumers and producers (see Figure 5). This pattern is inconsistent with increased competition driving the consumer-specific decline in the share of answers accepted.

\begin{figure*}
  \centering
\includegraphics[width=0.5\linewidth]{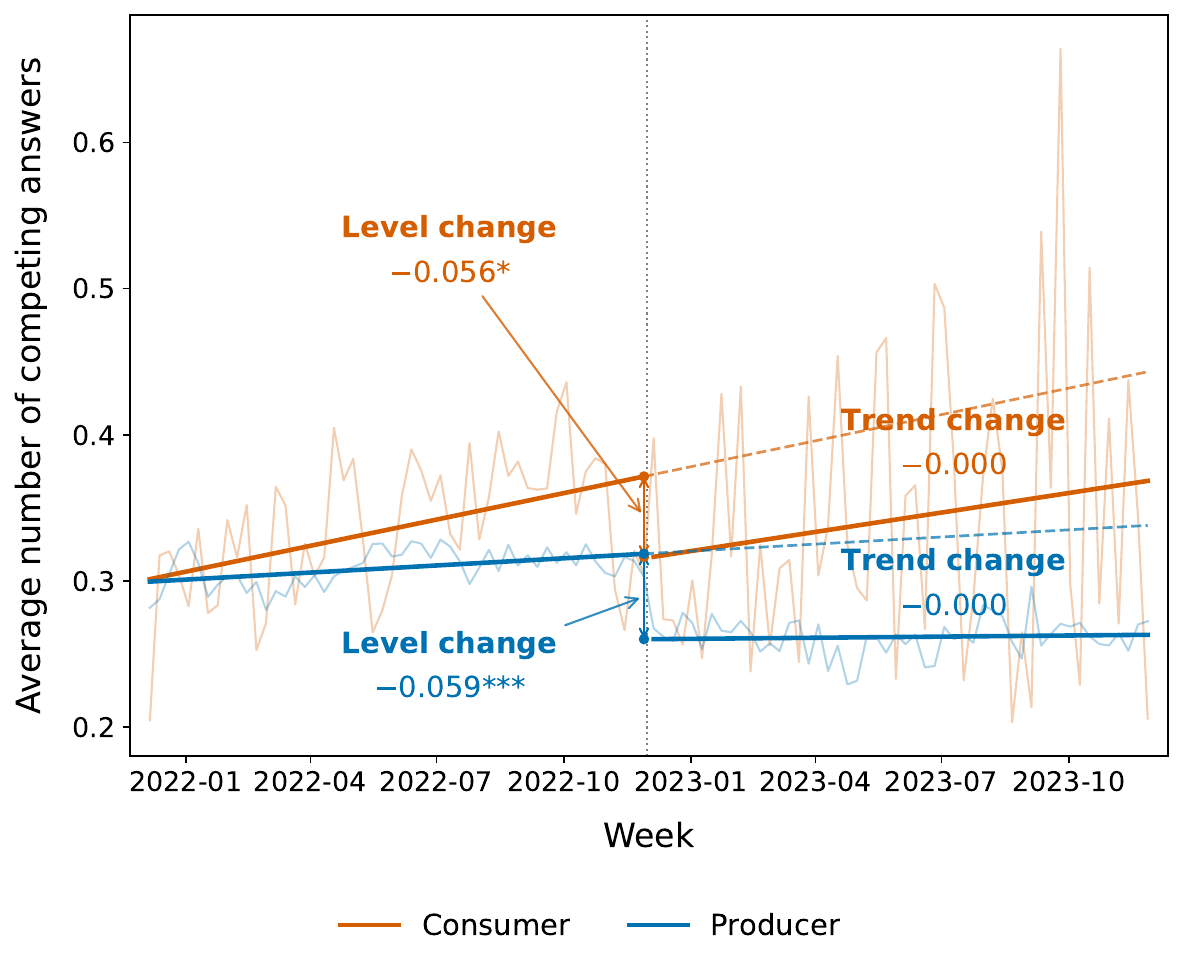}
  \caption{Weekly trends in the average number of competing answers already posted at the time of each response, shown separately for consumers and producers. The dotted vertical line marks the introduction of ChatGPT. Solid lines are fitted ITS trends; dashed lines extrapolate each group's pre-event trend forward as a counterfactual of what would have occurred absent the event. For each group, the annotated level change reports the immediate post-event shift relative to the counterfactual, while the annotated trend change reports the change in the weekly slope after the event relative to the pre-event slope.}
   \Description{Orange represents consumers and blue represents producers. Before ChatGPT, the fitted number of competing answers rises for both groups, more steeply for consumers: from about 0.30 to 0.37 for consumers and from 0.30 to 0.32 for producers. At the event, consumers show a statistically significant level decrease of 0.056 and producers a larger-significance decrease of 0.059, leaving fitted values near 0.32 and 0.26, respectively. Afterward, the consumer trend rises gradually to about 0.37 by late 2023 but remains below the counterfactual of roughly 0.44. The producer trend stays nearly flat near 0.26, compared with a counterfactual rising to about 0.34. The negative trend changes for both groups round to −0.000 and are not statistically significant.}
\end{figure*}

\section{Analysis Results for All Four User Groups}
\label{full_version}
This section presents the results of analyses using all four user groups: consumers, producers, hybrid users, and low-activity users. Consumers serve as the reference category in all models. Tables 3–5 report the results for average activity volume per user, average user–AI similarity, and average share of answers accepted, respectively.

\begin{table}
\centering
\caption{ITS analysis results for question and answer volumes, with the consumer group as the reference. Model 1 reports the results for the average number of questions per user, and Model 2 reports the results for the average number of answers per user. Standard errors are in parentheses. $^{*}p<0.05$; $^{**}p<0.01$; $^{***}p<0.001$.}
\label{tab_role_reversals}
\renewcommand{\arraystretch}{1.1}
\begin{tabular}{l l@{\hskip 5pt}l @{\hskip 15pt} l@{\hskip 5pt}l}
\toprule
 & \multicolumn{2}{c}{Model 1} & \multicolumn{2}{c}{Model 2} \\
 & \multicolumn{2}{c}{Question volume} & \multicolumn{2}{c}{Answer volume} \\
\midrule
Intercept & \hphantom{$-$}1.645$^{***}$ & (0.01) & \hphantom{$-$}1.100$^{***}$ & (0.01)\\
Hybrid & $-0.292$$^{***}$ & (0.01) & \hphantom{$-$}0.515$^{***}$ & (0.02)\\
Low-activity & $-0.565$$^{***}$ & (0.01) & \hphantom{$-$}0.080$^{***}$ & (0.01)\\
Producer & $-0.521$$^{***}$ & (0.01) & \hphantom{$-$}2.657$^{***}$ & (0.03)\\
\midrule
$Time$ (Pre-event trend) & $-0.002$$^{***}$ & (0.00) & $-0.000$ & (0.00)\\
Hybrid $\times$ $Time$ & \hphantom{$-$}0.001$^{*}$ & (0.00) & $-0.002$$^{**}$ & (0.00)\\
Low-activity $\times$ $Time$ & \hphantom{$-$}0.002$^{***}$ & (0.00) & $-0.000$ & (0.00)\\
Producer $\times$ $Time$ & \hphantom{$-$}0.001$^{**}$ & (0.00) & $-0.003$$^{**}$ & (0.00)\\
\midrule
$Post$ (Level change) & $-0.054$$^{**}$ & (0.02) & \hphantom{$-$}0.175$^{***}$ & (0.04)\\
Hybrid $\times$ $Post$ & \hphantom{$-$}0.008 & (0.02) & $-0.080$ & (0.05)\\
Low-activity $\times$ $Post$ & \hphantom{$-$}0.089$^{***}$ & (0.02) & $-0.024$ & (0.04)\\
Producer $\times$ $Post$ & \hphantom{$-$}0.081$^{***}$ & (0.02) & $-0.391$$^{***}$ & (0.07)\\
\midrule
$TimeAfter$ (Trend change) & $-0.001$$^{**}$ & (0.00) & $-0.001$ & (0.00)\\
Hybrid $\times$ $TimeAfter$ & \hphantom{$-$}0.001$^{**}$ & (0.00) & $-0.003$ & (0.00)\\
Low-activity $\times$ $TimeAfter$ & \hphantom{$-$}0.001 & (0.00) & \hphantom{$-$}0.001 & (0.00)\\
Producer $\times$ $TimeAfter$ & \hphantom{$-$}0.001 & (0.00) & $-0.012$$^{***}$ & (0.00)\\
\midrule
Adjusted $R^2$ & \multicolumn{2}{l}{\hspace{6pt}0.98} & \multicolumn{2}{l}{0.99}\\
\bottomrule
\end{tabular}
\Description{Before ChatGPT, consumers averaged 1.645 questions and 1.100 answers per user. Relative to consumers, all other groups posted fewer questions, while hybrids, low-activity users, and especially producers posted more answers; producers averaged an estimated 2.657 additional answers per user. Consumer question volume had a significant pre-event decline of 0.002 per week. At the event, consumers experienced a significant decrease of 0.054 questions per user and an increase of 0.175 answers per user. The producer differential implies a smaller increase of 0.027 questions and a decrease of 0.216 answers relative to its own counterfactual. Producers also had a significantly more negative post-event answer trend than consumers: combining the consumer trend change of −0.001 with the producer differential of −0.012 gives a producer trend change of −0.013 per week. For question volume, low-activity users had a significantly more positive level change than consumers, while hybrid and producer interactions offset much of the consumer decline. Hybrid and producer question-trend interactions were positive, whereas the low-activity interaction was not significant. The models explain 98\% of variation in question volume and 99\% in answer volume. Because the table tests interaction terms against the consumer reference, significance of a group’s combined change cannot be inferred directly from the stars on its interaction coefficient.
}
\end{table}

\begin{table}
\centering
\caption{ITS analysis results for the average user–AI similarity, with the consumer group as the reference. Each column reports estimates from a model using similarity scores generated by a different embedding model: OpenAI’s text-embedding model (Model 1) and Sentence Transformers (Model 2). Standard errors are in parentheses. $^{*}p<0.05$; $^{**}p<0.01$; $^{***}p<0.001$.}
\label{tab_ai}
\renewcommand{\arraystretch}{1.1}
\begin{tabular}{l l@{\hskip 5pt}l @{\hskip 15pt} l@{\hskip 5pt}l}
\toprule
 & \multicolumn{2}{c}{Model 1} & \multicolumn{2}{c}{Model 2} \\
 & \multicolumn{2}{c}{OpenAI} & \multicolumn{2}{c}{SBERT} \\
\midrule
Intercept & \hphantom{$-$}0.562$^{***}$ & (0.00) & \hphantom{$-$}0.568$^{***}$ & (0.00)\\
Hybrid & \hphantom{$-$}0.025$^{***}$ & (0.01) & \hphantom{$-$}0.014$^{*}$ & (0.01)\\
Low-activity & \hphantom{$-$}0.011$^{*}$ & (0.01) & \hphantom{$-$}0.006 & (0.00)\\
Producer & \hphantom{$-$}0.056$^{***}$ & (0.00) & \hphantom{$-$}0.036$^{***}$ & (0.00)\\
\midrule
$Time$ (Pre-event trend) & $-0.000$ & (0.00) & $-0.000$$^{*}$ & (0.00)\\
Hybrid $\times$ $Time$ & $-0.000$ & (0.00) & $-0.000$ & (0.00)\\
Low-activity $\times$ $Time$ & $-0.000$ & (0.00) & \hphantom{$-$}0.000 & (0.00)\\
Producer $\times$ $Time$ & \hphantom{$-$}0.000 & (0.00) & \hphantom{$-$}0.000 & (0.00)\\
\midrule
$Post$ (Level change) & \hphantom{$-$}0.018$^{***}$ & (0.00) & \hphantom{$-$}0.017$^{***}$ & (0.00)\\
Hybrid $\times$ $Post$ & \hphantom{$-$}$0.001$ & (0.01) & \hphantom{$-$}0.000 & (0.01)\\
Low-activity $\times$ $Post$ & $-0.001$ & (0.01) & $-0.003$ & (0.01)\\
Producer $\times$ $Post$ & $-0.019$$^{**}$ & (0.01) & $-0.018$$^{**}$ & (0.01)\\
\midrule
$TimeAfter$ (Trend change) & $-0.000$ & (0.00) & $-0.000$ & (0.00)\\
Hybrid $\times$ $TimeAfter$ & \hphantom{$-$}0.000 & (0.00) & \hphantom{$-$}0.001 & (0.00)\\
Low-activity $\times$ $TimeAfter$ & \hphantom{$-$}0.000 & (0.00) & \hphantom{$-$}0.000 & (0.00)\\
Producer $\times$ $TimeAfter$ & \hphantom{$-$}0.000 & (0.00) & \hphantom{$-$}0.000 & (0.00)\\
\midrule
Adjusted $R^2$ & \multicolumn{2}{l}{\hspace{6pt}0.73} & \multicolumn{2}{l}{0.25}\\
\bottomrule
\end{tabular}
\Description{Results are broadly consistent across embedding models. Before ChatGPT, consumers had estimated similarity scores of 0.562 with OpenAI embeddings and 0.568 with SBERT. Producers had the highest baseline similarity, exceeding consumers by 0.056 and 0.036, respectively; both differences are significant at p < 0.001. Hybrids also had significantly higher baseline similarity in both models, while the low-activity difference was significant only with OpenAI embeddings. Consumers show significant post-event level increases of 0.018 with OpenAI and 0.017 with SBERT. Producer-by-post interactions are negative and significant (−0.019 and −0.018), indicating that producers’ level changes were smaller than consumers’; combining coefficients yields an approximately −0.001 level change for producers in each model. Hybrid and low-activity post-event level changes do not differ significantly from the consumer change. Pre-event slopes, post-event trend changes, and group differences in those trend changes are all near zero, with little evidence of significant differences except for a small negative consumer pre-event trend in the SBERT model. Adjusted R-squared is 0.73 for the OpenAI model and 0.25 for SBERT. Stars on interaction terms test differences from consumers, not whether each group’s combined change differs from zero.}
\end{table}

\begin{table}
\centering
\caption{ITS analysis results for the average share of answers accepted, with the consumer group as the reference. Standard errors are in parentheses. $^{*}p<0.05$; $^{**}p<0.01$; $^{***}p<0.001$.}
\label{tab2}
\renewcommand{\arraystretch}{1.1}
\begin{tabular}{l l@{\hskip 5pt}l}
\toprule
 & \multicolumn{2}{c}{Share of answers accepted} \\
\midrule
Intercept & \hphantom{$-$}0.343$^{***}$ & (0.01)\\
Hybrid & $-0.049$$^{***}$ & (0.01)\\
Low-activity & $-0.169$$^{***}$ & (0.01)\\
Producer & \hphantom{$-$}0.058$^{***}$ & (0.01)\\
\midrule
$Time$ (Pre-event trend) & \hphantom{$-$}0.000 & (0.00)\\
Hybrid $\times$ $Time$ & $-0.000$ & (0.00)\\
Low-activity $\times$ $Time$ & $-0.000$ & (0.00)\\
Producer $\times$ $Time$ & $-0.000$$^{*}$ & (0.00)\\
\midrule
$Post$ (Level change) & $-0.020$$^{*}$ & (0.01)\\
Hybrid $\times$ $Post$ & \hphantom{$-$}0.005 & (0.01)\\
Low-activity $\times$ $Post$ & \hphantom{$-$}0.073$^{***}$ & (0.01)\\
Producer $\times$ $Post$ & \hphantom{$-$}0.043$^{***}$ & (0.01)\\
\midrule
$TimeAfter$ (Trend change) & $-0.001$$^{**}$ & (0.00)\\
Hybrid $\times$ $TimeAfter$ & \hphantom{$-$}0.000 & (0.00)\\
Low-activity $\times$ $TimeAfter$ & \hphantom{$-$}0.000 & (0.00)\\
Producer $\times$ $TimeAfter$ & \hphantom{$-$}0.001 & (0.00)\\
\midrule
Adjusted $R^2$ & \multicolumn{2}{l}{\hspace{6pt}0.94}\\
\bottomrule
\end{tabular}
\Description{Before ChatGPT, consumers had an estimated acceptance share of 0.343. Relative to consumers, the baseline share was 0.049 lower for hybrids and 0.169 lower for low-activity users, but 0.058 higher for producers; all three differences are significant at p < 0.001. Pre-event trends were close to zero, although producers had a small but significant more-negative slope than consumers. At the event, consumers experienced a significant level decrease of 0.020. The corresponding interaction was not significant for hybrids, but was positive and significant for low-activity users (0.073) and producers (0.043), indicating level changes more positive than the consumer change. Combining coefficients gives estimated level changes of approximately −0.015 for hybrids, +0.053 for low-activity users, and +0.023 for producers. Consumers also experienced a significant post-event trend change of −0.001. None of the group interactions with this trend change were statistically significant, indicating no clear evidence that their trend changes differed from that of consumers. The model has an adjusted R-squared of 0.94. Significance markers on interaction coefficients test differences from consumers; they do not test whether each group’s combined level or trend change differs from zero.}
\end{table}

\end{document}